\documentclass{article}

\usepackage[english]{babel}

\usepackage[letterpaper,top=2cm,bottom=2cm,left=3cm,right=3cm,marginparwidth=1.75cm]{geometry}

\usepackage{amsmath}
\usepackage{graphicx}
\usepackage[colorlinks=true, allcolors=blue]{hyperref}

\usepackage{subfigure}
\usepackage{float}
\usepackage{amsmath}
\usepackage{datetime}
\usepackage{indentfirst}

\title{Exploring the Endogenous Nature of Meme Stocks Using the Log-Periodic Power Law Model and Confidence Indicator}
\author{Hideyuki Takagi}
\date{September 2021}

\begin{document}
\maketitle

\begin{abstract}
\noindent This study examined the endogenous nature of negative bubbles forming in meme stocks with the Log-Periodic Power Law (LPPL) Confidence Indicator (CI). A meme stock is a stock that has gained a significant amount of attention on a large social media platform such as Yahoo! or Reddit. This study examined four meme stocks including Tesla, Inc. (TSLA), GameStop Corp. (GME), Koss Corporation (KOSS), and AMC Entertainment Holdings Inc (AMC). The CI was able to detect numerous bubbles forming in meme stocks, but had difficulty in significantly predicting social media-induced exogenous rallies. This may have been due to price movements affected by external causes such as short squeezes. However, the model did provide proof for the formation of previous bubbles that could have been a catalyst for the meme stocks rallies. This study outlines the real unpredictability of many black-swan events, and further studies could be done examining exogenous bubbles.
\end{abstract}

\section{Keywords}
\noindent Market Crash; Financial Bubble; Endogenous nature; Log-Periodic Power Law Model; Meme Stock
\section{Introduction}

Recently, there has been a rise in commission-free stockbrokers and an increase in retail investors easily purchasing securities. With the combination of accessible social media platforms, such as Reddit, Twitter and Discord, pockets of interest groups have begun to impact many investors and the stock market as a result. For example, members of the Reddit group r/WallStreetBets have engaged in financial discussions in which they have sometimes created entire movements around single stocks, often with speculative reasons for doing so \cite{[1]costola2021mementum}. Consequently, the GameStop Corporation observed a massive short squeeze triggered by such a movement in January 2021; the price of its stock was pushed to a staggering USD347.00 per share from around USD5.00 per share just a year before \cite{[2]o'mahony_moran_fool_adams_2020}. Many such stocks are labeled meme stocks; they are characterized by a belief in the potential of the company among those on social media platforms \cite{[2]o'mahony_moran_fool_adams_2020}. What makes these stocks special is that, unlike the tulip mania and dot-com bubbles, many of the speculative gains in meme stocks are non-traditional \cite{[1]costola2021mementum}. The phenomenon was began by a group of retail investors who openly coordinated stock decisions, and often, such group decisions were fundamentally different from decisions by institutional investors, such as hedge and mutual fund investors \cite{[1]costola2021mementum}.

This study analyzed the endogenous nature of a speculative bubble that many of these meme stocks experience by using the log-periodic power law (LPPL) model devised by Sornette et al \cite{[3]johansen2000crashes}. The model was initially created to detect positive bubbles or a “big crash” that plagues markets on occasion \cite{[3]johansen2000crashes}. Born from the phase transition idea in statistical physics, the model assumes that such crashes result from the collective behavior of “noise” traders imitating each other against “rational” traders. The noise traders being the ones that act with decisions that they believe is helpful without any proper information \cite{[4]scott_2021}. The LPPL model formulates the potential for a crash by observing faster than exponential growth in an asset’s price and by observing accelerating log-periodic volatility fluctuations due to a “battle” between buyers and sellers. It has sufficiently captured many bubbles, such as the crash of the U.S. Stock Market in 2020 \cite{[5]SHU2021101497}, Bitcoin crashes \cite{[6]wheatley2019bitcoin}, and Shanghai Stock Exchange crashes \cite{[7]shu2020detection}. 

The LPPL model have been extensively used to study different types of bubbles. Sornette et al \cite{[3]johansen2000crashes} first proposed then illustrated how the model could be fitted onto the S\&P 500 and Dow Jones historical data. Filimonov and Sornette altered the LPPL model by reducing the non-linear parameters to make calibration easier \cite{[19]filimonov2013stable}. Sornette et al \cite{[11]sornette2015real} proposed the LPPLS Confidence Indicator (CI) to analyze the real-life bubble crash of the 2015 Shanghai Stock Market. Wheatley et al \cite{[6]wheatley2019bitcoin} used the LPPLS model to analyze bitcoin bubbles. Shu et al \cite{[5]SHU2021101497} utilized the CI to examine the 2020 Covid crash of the US market as well as the detection of Chinese stock market bubbles \cite{[7]shu2020detection}. This paper adds to the study of rebound bubbles by looking at the opposite of crashes––negative bubbles.

A negative bubble is the opposite of a positive bubble; it often shows critical points as lows, and “bursts” of the bubble as rallies or sideways movements \cite{[8]demirer2019predictability}. This study examined the significantly high sell positions that formulate negative bubbles and the nature of meme stocks as “bursts” of related rallies initiated by speculating retail investors. This research addressed the recent phenomenon of these black-swan rallies and their characteristics. This was done by employing the LPPL model using a confidence indicator (CI), which is discussed in the methodology section. To this end, four of the most social media mentioned \cite{[23]guimaraes2021short} thus influential meme stocks were examined: TSLA, GME, KOSS and AMC. They were taken from memestocks.org, a website that displays meme stocks mentioned the most in the r/WallStreetBets subreddit \cite{[9]davis_2021}. Daily data was collected through Yahoo! Finance \cite{[10]yahoo_2021}.

The LPPL model detects endogenously created or internally-causing bubbles rather than exogenous ones. The abnormal bubble created by the meme stock phenomenon appeared to be exogenous. However, this study utilized the model to examine the log-periodic nature of prices to determine if the rally truly was exogenous \cite{[11]sornette2015real}, as it was possible that the uncommon price movements were formulated by an existing endogenous negative bubble movement. With uncertainty plaguing the markets, this paper shows how prices can occasionally be driven by completely arbitrary factors that are out of sight. Section 3 presents the methodology that was utilized for this research, including the LPPL model, its calibration and the CI. Section 4 discusses the results, and Section 5 presents this study’s conclusions.

\section{Methods}

\subsection{The Log-Periodic Power Law Model (LPPL)}

The Log-Periodic Power Law Model (LPPL) was first devised by Sornette et al \cite{[3]johansen2000crashes}. The model incorporates assumptions of the rational expectation of traders, the herding and imitation of traders, and the use of the diffusion model concept in statistical physics and mathematics \cite{[12]srivastava2017martingale}. This fusion of concepts characterizes the bubble by a faster-than-exponential growth of price deviating from its fundamental value as well as accelerating log-periodic fluctuations near the crash point \cite{[3]johansen2000crashes}. The dynamics of the price in the LPPL model follows as:

\begin{equation}
dp=\mu{(t)}p(t)dt+\sigma{(t)}p(t)dW-kp(t)dj
\end{equation}

Where $\mu{(t)}$ is the expected return, $\sigma{(t)}$ is the volatility, $dW$ is the increment of a Wiener process \cite{[12]srivastava2017martingale}, $dj$ defining a jump process with value 0 before and 1 after the crash, and $k$ being the amplitude of a potential crash. The concept of the crash hazard rate $h(t)$, is defined as the dynamics of the jump process and is calculated as $E[dj]=h(t)dt$. The term also determines the growth of the price as agents imitate each other \cite{[3]johansen2000crashes}. Thus, determining the hazard rate is crucial in finding the critical time.
\bigskip

With rational expectation conditions, the unconditional expectation of the price should follow as $E[dp]=0$. Thus,

\begin{equation}
E[dp]=kh(t)p(t)dt 
\end{equation}

The LPPL model assumes that there are rational traders as well as irrational noise traders. The latter is characterized by its collective herding behavior as traders imitate each other. The crash occurs when enough traders imitate and thus cause a massive sell-off. Borrowing the concept of stochastic dynamical model of interacting particles \cite{[13]liggett2012interacting}, Sornette et al \cite{[3]johansen2000crashes} describes the effect of the noise traders during a bubble with the hazard rate of crash as:

\begin{equation}
h(t)\approx B_{0}(t_{c}-t)^{-\alpha}+B_{1}(t_{c}-t)^{-\alpha}\cos{[\omega ln(t_{c}-t)+\psi']}
\end{equation}

One important characteristic of this model is the increase in accelerating oscillation as it reaches a critical time following the log-periodic frequency of $\frac{\omega}{2\pi}$ \cite{[14]geraskin2013everything,[15]derrida1983fractal}. Solving equation (2) with equation (3), Sorenette et al \cite{[3]johansen2000crashes} shows the LPPL model as:

\begin{equation}
    ln[p(t)]\approx A+B(t_{c}-t)^{\beta}[1+C\cos(\omega ln(t_{c}-t)+\phi)]
\end{equation}

where $A >0$ is $E[ln[p(t)]]$ at the critical time, $-\frac{k\alpha\beta}{\sqrt{m^{2}+\omega^{2}}}$ is the magnitude of oscillations around the critical time and $B$ is the decrease in $ln[p(t)]$ over the time unit if $C$ is close to zero. $0<\beta<1$ is the exponent of the PL growth, $\omega$ is the log-frequency of the oscillations in the bubble and $0<\phi<2\pi$ is a phase transition parameter \cite{[14]geraskin2013everything}. Equation (4) is therefore the LPPL model devised by Sornette et al \cite{[3]johansen2000crashes}. and describes the price before a potential crash. There are two main features of the model: a faster than exponential growth in the price of the asset before a crash and an increasing number of oscillations as the price approaches the critical time \cite{[16]sornette2003critical}.

\subsection{Positive and Negative Bubbles}
In the context of the LPPL model, there are two main types of bubbles: positive and negative. The former is a traditional bubble characterized by an increasing price with a dramatic fall once a critical time is reached. The latter is a mirror image of a positive bubble characterized by a downward accelerating price with a rebound or bull run of the price once a critical time is reached \cite{[17]yan2010diagnosis}. An example of a positive bubble can be seen in Figure (a), and an example of a negative bubble can be seen in Figure (b) \cite{[18]chair_2019}. In the LPPL model, the difference between positive and negative bubbles is the $B$ coefficient. $B$ is less than zero for a positive bubble, while $B$ is greater than zero for a negative bubble \cite{[3]johansen2000crashes}.

\begin{figure}[H]
\centering
\mbox{\subfigure[A positive bubble from the Financial Crash Observatory (FCO) presentation \cite{[18]chair_2019}.]{\includegraphics[width=3in]{Figure1.pdf}}
\quad
\subfigure[A positive bubble from the FCO presentation \cite{[18]chair_2019}.]{\includegraphics[width=3in]{Figure2.pdf}}}
\label{fig12}
\end{figure}

\subsection{Calibration}
The LPPL model has seven different parameters, so optimizing the model for a specific dataset is an arduous task \cite{[19]filimonov2013stable}. Many calibration techniques rely on slaving three linear parameters to four non-linear parameters, making a very complex quasi-periodic structure with multiple minima \cite{[19]filimonov2013stable}. This study utilized a calibration method suggested by Filiminov and Sornette; it represents the LPPL model with four linear and three non-linear parameters, and this significantly reduces the complexity \cite{[19]filimonov2013stable}. This is first done by rewriting the LPPL model by expanding the cosine term and introducing two new parameters: $C_{1}=C\cos\phi$ and $C_{2}=C\sin\phi$ \cite{[7]shu2020detection,[19]filimonov2013stable}. This makes the new LPPL representation:

\begin{equation}
    ln[Ep(t)]\approx A+B(t_{c}-t)^{\beta}+C_{1}(t_{c}-t)^{\beta}\cos(\omega ln(t_{c}-t))+C_{2}(t_{c}-t)^{\beta}\sin(\omega ln(t_{c}-t))
\end{equation}

The function has three non-linear $(t_{c},\omega, \beta)$ and four linear $(A,B,C_{1},C_{2})$ parameters. Using the least-squares method,

\begin{equation}
    F(t_{c},\beta,\omega,A,B,C_{1},C_{2})=\sum_{i=1}^{N} [lnp(\tau_{i})-A-B(t_{c}-\tau_{i})^{\beta}-C_{1}(t_{c}-\tau_{i})^{\beta}\cos(\omega ln(t_{c}-\tau_{i})^\beta\sin(\omega ln(t_{c}-\tau_{i}))]^{2}
\end{equation}

Slaving the four linear parameters to the three non-linear ones, one gets the non-linear optimization problem:

\begin{equation}
    (\hat{\tau_{c}},\hat{\beta},\hat{\omega})=arg\min_{(t_{c},\beta,\omega)}F_{1}(t_{c},\beta,\omega)
\end{equation}

with the cost function $F(x)$ given using:

\begin{equation}
    F_{1}(t_{c},\beta,\omega)=\min_{(A,B,C_{1},C_{2})}F(t_{c},\beta,\omega,A,B,C_{1},C_{2})
\end{equation}

The linear parameters have the following solution in the form of a matrix equation \cite{[19]filimonov2013stable}.

$$
\begin{bmatrix}
N & \sum f_{i} & \sum g_{i} & \sum h_{i} \\
\sum f_{i} & \sum f_{i}^{2} & \sum f_{i} g_{i} & \sum f_{i} h_{i} \\
\sum g_{i} & \sum f_{i} g_{i} & \sum g_{i}^{2} & \sum h_{i} g_{i} \\
\sum h_{i} & \sum f_{i} h_{i} & \sum g_{i} h_{i} & \sum h_{i}^{2} 
\end{bmatrix}  
\begin{bmatrix}
\hat{A} \\
\hat{B} \\
\hat{C_{1}} \\
\hat{C_{2}}
\end{bmatrix} 
= 
\begin{bmatrix}
\sum lnp_{i} \\
\sum f_{i}lnp_{i} \\
\sum g_{i}lnp_{i} \\
\sum h_{i}lnp_{i}  
\end{bmatrix} 
$$

The change in the LPPL model means that the complexity is significantly reduced, and the reduction of minima translates to the possible use of rigorous search methods. This study utilized the Nelder-Mead simplex method to search for the minima, as suggested by Filiminov and Sornette \cite{[19]filimonov2013stable}. To this end, an algorithmic code devised by Joshua Nielsen \cite{[20]Nielsen_2020} and based on the method expressed by Filiminov and Sornette \cite{[19]filimonov2013stable} was used that can be accessed in the references section below. 

\subsection{The LPPL CI}

The LPPL CI, first introduced by Sornette et al \cite{[8]demirer2019predictability}, is defined as “the fraction of fitting window in which LPPL calibrations satisfy specified filter conditions” \cite{[7]shu2020detection}. A spike in the indicator’s value suggests the possibility of a bubble as it suggests that the LPPL pattern can be seen in multiple time scales \cite{[11]sornette2015real}. Small values for the indicator signal fragility as they mean the indicator may not capture the LPPL model in sufficient time windows \cite{[11]sornette2015real}. 

The CI is calculated by fitting specified time window $dt$ from the beginning of dataset $t_{1}$ until the end of the dataset $t_2$ \cite{[11]sornette2015real}. Then the model is gauged and the number of specified frames that the dataset fits with respect to filter constraints is counted. The number is divided by the total number of fitting windows, $dt$. Nielsen’s code included in the indicator for both positive and negative bubbles \cite{[20]Nielsen_2020}.  The calibration conditions follow Sornette et al.’s recommendation to prevent data overfitting by altering conditions \cite{[1]costola2021mementum}.  The algorithms and conditions were standardized for all the datasets used in this study. 

\section{Results and Discussion}

Numerous studies that have used the LPPL model have examined positive “traditional” crashes (e.g., a study by Brée and Joseph \cite{[21]bree2013testing}), but this study researched negative, non-traditional crashes by examining TSLA, GME, KOSS and AMC datasets exported from Yahoo! Finance \cite{[10]yahoo_2021} using the code made by Nielsen \cite{[20]Nielsen_2020}. Most of the stocks experienced enormous gains in price that followed oscillating downward prices. CIs indicated local minima for the stocks’ prices in many cases. Rallies were then forecasted using a negative bubble indicator. Though the study focused on bull runs, positive bubbles were also shown in the results as Nielsen’s code also calculated them.

\begin{figure}[H]
  \includegraphics[width=\linewidth]{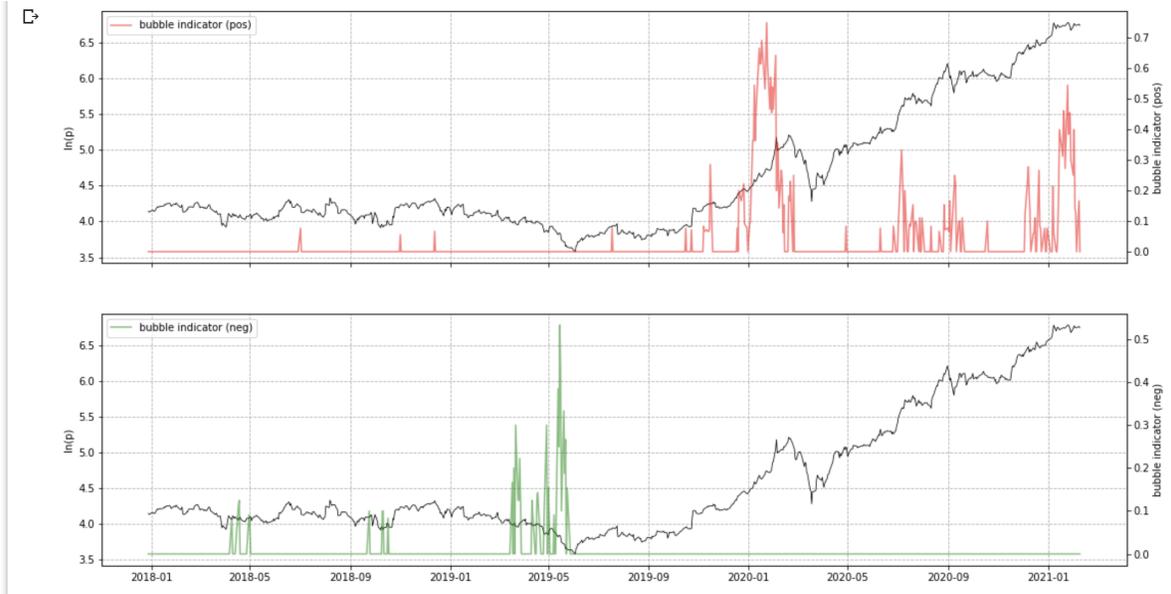}
  \caption{The TSLA stock price from July 11th, 2017 to February 9th, 2021. The negative LPPL CI is shown in green, while the positive LPPL CI is shown in red.}
\end{figure}

\begin{figure}[H]
  \includegraphics[width=\linewidth]{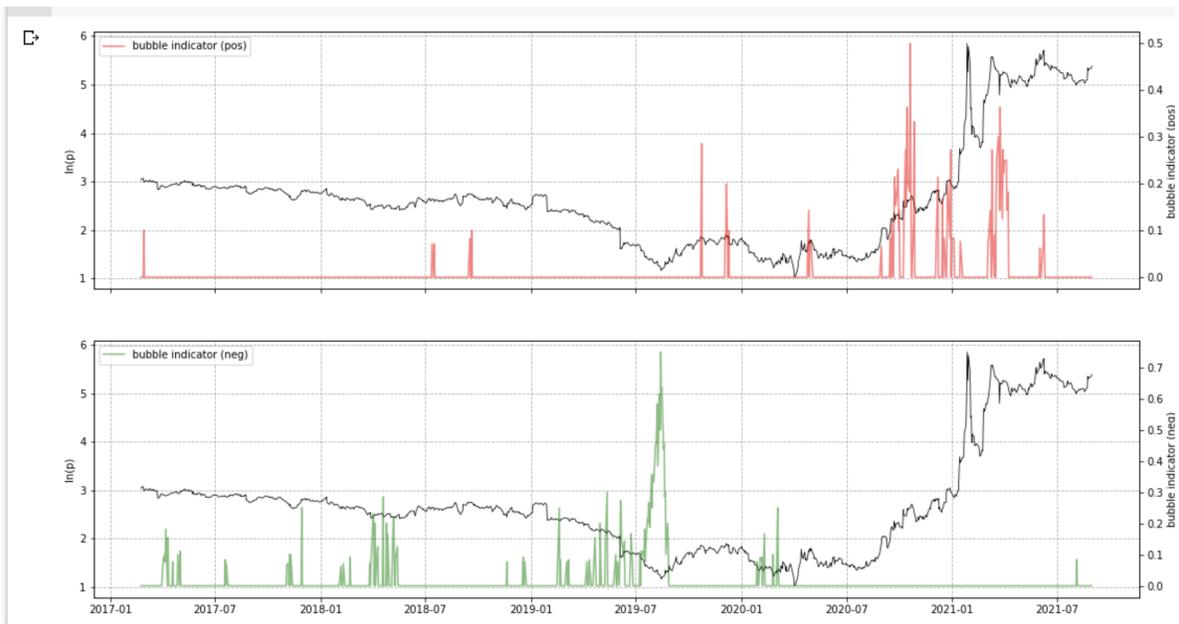}
  \caption{The GME stock price from September 1st, 2016 to August 31st, 2021. The negative LPPL CI is shown in green, while the positive LPPL CI is shown in red.}
\end{figure}

\begin{figure}[H]
  \includegraphics[width=\linewidth]{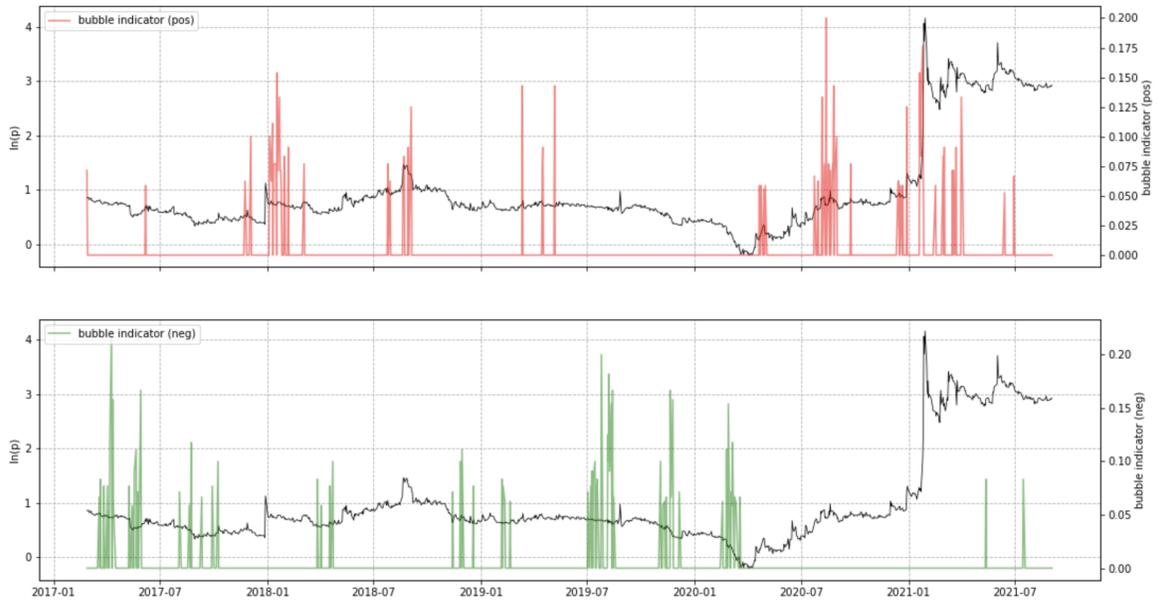}
  \caption{The KOSS stock price from September 6th, 2016 to September 3rd, 2021. The negative LPPL CI is shown in green, while the positive LPPL CI is shown in red.}
\end{figure}

\begin{figure}[H]
  \includegraphics[width=\linewidth]{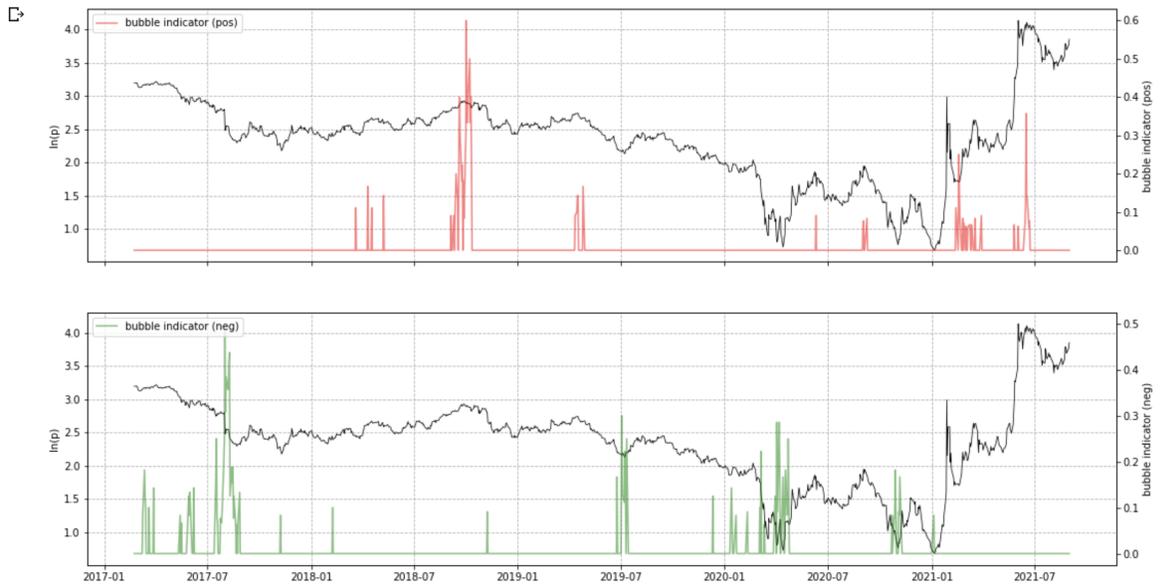}
  \caption{The AMC stock price from September 1st, 2016 to August 31st, 2021. The negative LPPL CI is shown in green, while the positive LPPL CI is shown in red.}
\end{figure}

In Figures 1–4, as aforementioned, red denotes positive bubble indicators and green denotes negative bubble indicators. All the datasets have more than 900 data points; Demirer et al. has noted that long-term bubble indicators should have more than 500 data points to ensure a study’s efficacy \cite{[8]demirer2019predictability}. A summary of the rally statistics and related information is in Table 1.

\begin{table}[H]
\caption{Negative bubble statistics for the meme stocks, detected using the LPPL CI and based on daily data.}
\label{tab:my-table}
\begin{tabular}{|c|c|c|c|c|c|c|}
\hline
Stock & Bottom Day & Bottom Price & Rally Date & Rally Price & Rally Size & CI Value \\ \hline
TSLA & June 3rd, 2019 & 35.7 & February 20th, 2020 & 180 & 404\% & $\sim$0.58 \\ \hline
GME & August 15th, 2019 & 3.21 & December 6th, 2019 & 6.68 & 108\% & $\sim$0.8 \\ \hline
KOSS & April 3rd, 2020 & 0.82 & January 29th, 2021 & 64 & 7705\% & $\sim$0.15 \\ \hline
AMC & April 13th, 2020 & 2.08 & June 8th, 2020 & 6.45 & 210\% & $\sim$0.27 \\ \hline
\end{tabular}
\end{table}

It should be noted that the indicators showed the formation of potential bubbles spikes more often than they showed actual bubbles. Thus, they were not accurate in that they showed only when bubbles formed, but they were accurate in that they showed potential bubbles \cite{[14]geraskin2013everything}. This study therefore focused on the indicator values that accurately predicted bubbles. This paper’s interpretation of the data can be seen, for example, in Figure 4. The figure shows multiple signals for negative bubbles; the relevant signals are indicator values greater than or equal to 0.3. 

This study found a spike in the TSLA stock price right after a negative bubble indicator showed an uptick of about 0.58 with a bull run from June 2019 to a pullback in March 2020. The indicator failed to capture a rally that began after March 2020. However, entering into trades right after the negative bubble indicator was a sign that would have generated enormous profit of 404\% for investors (Table 1).

For the stock GME, the highest negative bubble indicator peak showed a value of about 0.8 in August 2019, which was a strong signal that a bounce-back in price was about to occur. However, the LPPL indicator failed to capture an enormous rally that began in the second half of 2020. The rally was potentially fueled by an influx of retail investors in meme stocks, as suggested by Costola et al \cite{[1]costola2021mementum}. Thus, it was exogenous in nature, which the LPPL indicator could not capture. One could argue that an indicator value of about 0.25 in March 2020 implied a negative bubble, but it may have been only a fragility that resulted from the model as mentioned by Sornette et al \cite{[11]sornette2015real}.

The negative bubble indicator did capture a KOSS rally that started at the bottom in April 2020. A cluster of indicator values formed around that time, wherein local minima formed and the highest indicator value was about 0.15. However, the study also found two other indicator clusters that formed around August 2019 and December 2019, respectively, but no signs of actual negative bubbles forming. In addition, a positive bubble indicator that spiked in January 2021 suggested a crash but was instead met with a significant increase in price over a short period of time. This could have been due to the fact that the LPPL model captures endogenous bubbles and the spike in January 2021 had an exogenous origin \cite{[22]johansen2002endogenous} due to a short squeeze catalyzed by an increase in social media herding \cite{[1]costola2021mementum}.

Figure 4 shows negative bubble indicator spikes in August 2017, July 2019, April 2020 and October 2020, with values of about 0.5, 0.27, 0.27 and 0.2, respectively. The highest indicator value was about 0.5 in August 2017, but the price did not follow a rally. Moreover, a significant spike that followed in January 2021 was not captured by the LPPL model. Again, this may have been due to an exogenous influence on the AMC stock price as a result of the emergence of the meme stock phenomenon. An indicator of about 0.05 right before the spike may have been only the fragility of the signal \cite{[11]sornette2015real}.

As aforementioned, Table 1 shows the statistics for the negative meme stock bubbles detected using the CI. The rallies shown in the table are not necessarily the largest, nor did they precede high CI values. Thus, this study examined rallies that were sufficiently indicated by a CI beforehand. For example, the stock AMC showed a large spike in its CI value around August 2017 but did not indicate a large bubble that formed in January 2021 due to a low CI value. The table shows a rally that followed in April 2020; it is indicated by a CI value of about 0.27. The rally price was the highest price that was achieved in that specific local rally. This study therefore concluded a rally by affirming a pullback in price after the rally’s price. Hence, the GME rally, for example, was concluded in December 2019. 

Group agreements among traders in social media organizations may have resembled agents imitating each other through a network, causing massive upticks in price (i.e., negative bubbles). Those using social media platforms may have increased the susceptibility of agents by creating a “fear of missing out” context, creating demand for specific stocks and pushing prices upwards. Meanwhile, hedge fund and other institutional investors may have seen these meme stocks as losses and fought to keep prices down by short-selling the stocks, thereby causing oscillations in price. The CIs may have not shown this clearly as the rallies may have been caused by short squeezes, which were triggered when retail traders engaged in simultaneous buying \cite{[23]guimaraes2021short}.

\section{Conclusion}

This study examined sudden rallies experienced by many meme stocks; this was done by employing the LPPL model and its indicators. The model was run using several different daily datasets for four different stocks, namely TSLA, GME, KOSS and AMC. Many of the meme stocks observed weekly gains of more than 50\%, as shown in Table 1. The purpose of examining the stocks was to consider the endogenous natures of such gains, which could be predicted using the LPPL bubble model. The results indicated that there were some endogenous signals several months before the actual rallies the stocks experienced. For example, by running the LPPL Confidence Indicator with the daily adjusted returns of the stock GME, a spike at around August 2019 with a CI value of about 0.8 was found. This indicated a clear negative LPPL bubble pattern that formed as a result of a higher than exponential decline in the stock’s price. However, an actual “big rally” was barely captured. This may have been due to the fact that the unnatural increase in prices stemmed from exogenous factors. 

Indeed, the LPPL model sufficiently predicted when bubbles formed months in advance and thereby signaled rallies, but market instability showed that such rallies may have had a completely different cause: short squeezes. The prediction of meme stock bubbles is a difficult task given exogenous factors, even if initial “causes” of bubbles are endogenous. This uncertainty outlines variability of the market as prices are occasionally driven by completely arbitrary factors. Further studies could be conducted to examine the exogenous natures of bubbles in more detail using the LPPL model or otherwise.

\bibliographystyle{unsrt}
\bibliography{bibliography}

\end{document}